# Rheumatoid Arthritis: Automated Scoring of Radiographic Joint Damage

Yan Ming Tan[1,#], Raphael Quek Hao Chong[2,#], Carol Anne Hargreaves[1]

[1] Department of Statistics and Data Science, National University of Singapore;
[2] Department of Electrical & Computer Engineering, National University of Singapore, Singapore;

## Abstract

Rheumatoid arthritis is an autoimmune disease that causes joint damage due to inflammation in the soft tissue lining the joints known as the synovium. It is vital to identify joint damage as soon as possible to provide necessary treatment early and prevent further damage to the bone structures. Radiographs are often used to assess the extent of the joint damage. Currently, the scoring of joint damage from the radiograph takes expertise, effort, and time. Joint damage associated with rheumatoid arthritis is also not quantitated in clinical practice and subjective descriptors are used. In this work, we describe a pipeline of deep learning models to automatically identify and score rheumatoid arthritic joint damage from a radiographic image. Our automatic tool was shown to produce scores with extremely high balanced accuracy within a couple of minutes and utilizing this would remove the subjectivity of the scores between human reviewers.

## Graphical Abstract

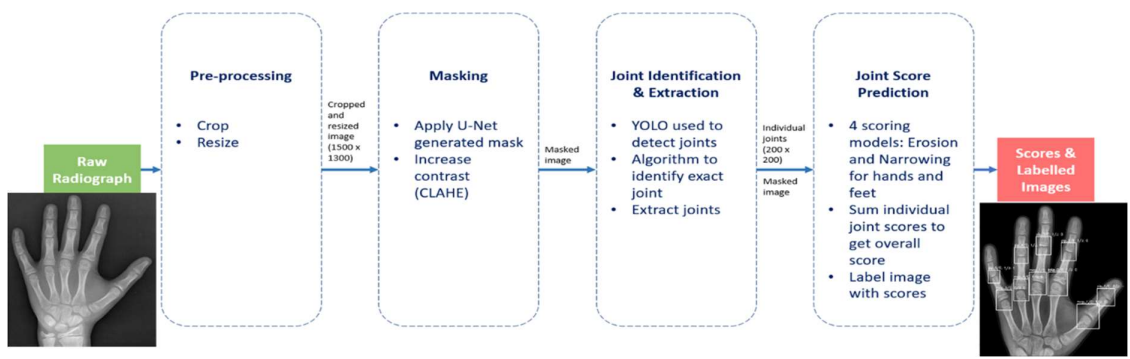

## 1. Introduction

Rheumatoid arthritis (RA) is an autoimmune disease where the immune system mistakenly attacks the body's own tissues. This causes inflammation in the synovium, which eventually leads to joint damage. External symptoms can include red and swollen joints accompanied by pain. About 0.5 - 1% of the global population are affected by RA [1]. Inflammation of the joint will slowly cause cartilage, the layer of tissue that covers the ends of the bones, to erode. As the amount of cartilage decreases, the joint space also narrows. Long-term inflammation can also cause an increase in osteoclasts, cells that break down the tissue in bones, resulting in bone erosion. The degrees of narrowing and erosion observed in radiographs for RA are used in the Sharp/van der Heijde (SvH) method [2] to measure joint damage. This method looks at specific joints in the hands and feet, usually linked to inflammation caused by RA. Radiographs can provide a fair representation of joint damage but are presently not used to their full potential because there is no fast way to measure the damage quantitatively [1]. Currently, the scoring of the degree of joint damage in RA patients is done by manually reviewing their radiographs. This is generally expensive as it takes effort and time. Additionally, joint damage associated with RA is not quantitated in clinical practice, but instead, subjective descriptors such as "mild, moderate, or severe" are used in official reports [1]. Thus, it is desirable to have a method that can quickly and objectively classify joints to allow for more consistent and accurate scoring in the clinical and research settings without the need for much medical expertise.

For image classification, Deep Learning (DL) models have been outperforming classical Machine Learning (ML) models since the 2012 ImageNet challenge [3]. The DL model, AlexNet, which contained 5 layers of convolutional neural network (CNN) had achieved 15.2% Top-5 classification error [4]. Thereafter, subsequent years of ImageNet challenges have been dominated by CNN DL models. CNN DL models can achieve such remarkable performance due to their abilities to extract essential features during training [5]. Raw images can be used directly as inputs without the need for prior feature extraction. The accuracies of these CNN DL models have been constantly increasing in plenty of diverse applications in computation vision medical tasks such as disease classification, brain cancer classification, organ segmentation, haemorrhage detection, and tumour detection [6]. Much work has also been done on applying CNN models to classifying X-ray images [7] [8], and how to enhance the image contrast [9]. Existing work done on the automatic scoring of erosion due to RA produced a model based on VGG16, a CNN architecture [10], that is as accurate as human scorers [11]. Instead of SvH, the

Ratingen erosion scoring was used [12]. The segmentation of the joints was not part of their work as they used pre-extracted joint images.

Our work built upon their approach and managed to achieve an even higher accuracy. As a deviation from their work, we treated the problem as a classification of ordinal classes by utilizing ordinal class encoding. To deal with the imbalanced data, an under-sampling approach was attempted. We additionally included the automatic segmentation and extraction of joints. State-of-the-art DL models for computer vision were applied to remove image noise, and accurately segment the joints, which ensured quality of the training samples. A lightweight U-Net architecture which was constructed for bones segmentation [13] was used for the purpose of removing background noise. As for robust joint detection in X-ray images, the YOLOv3 model [14] was implemented. This paper presents the results of our CNN model trained on these joint images for the automatic measurement of joint damage according to the SvH scoring method.

## 2. Materials and Methods

*2.1. Dataset*

The X-ray datasets used for the analyses described in this paper were contributed by the University of Alabama at Birmingham, and were compiled from two sources - CLEAR (Consortium for the Longitudinal Evaluation of African Americans with Rheumatoid Arthritis [15]), and TETRAD (Treatment Efficacy and Toxicity in RA Database and Repository: A study of RA patients starting biologic drugs [16]).

A total of 367 sets of 4 radiographs each per patient were provided as JPG files. Each patient had a radiograph of their left hand (LH), right hand (RH), left foot (LF), and right foot (RF). Corresponding SvH scores were given in CSV format. It includes each patient's overall total damage scores, total erosion scores, total narrowing scores, and the narrowing and erosion scores of each joint. In addition, it was noticed that the size of the X-ray images varies considerably. Thus, this called for the need to resize them before the images could be used for training *inputs*. In addition, the dataset consists of a large number and percentage of score = 0 for both narrowing and erosion (Fig. *1*).

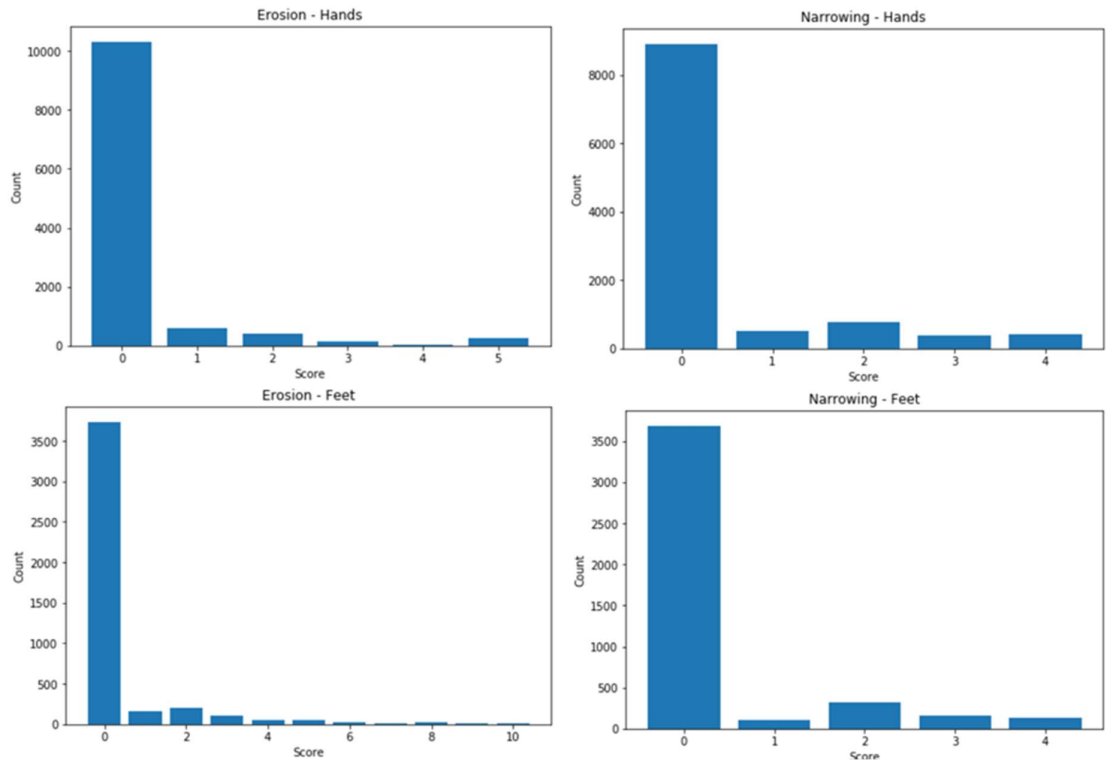

**Fig. 1.** Distributions of scores for erosion and narrowing of each limb.

## 2.2. Data Pre-processing

### 2.2.1. Normalization

All images were normalized to have pixel values ranging from 0 to 1 by dividing the image arrays by 255 before training as the original pixel values belong to {0, 1 …, 255}.

### 2.2.2. Image re-scaling and padding

All the images were resized to 1500 x 1200 pixels. The original aspect ratio was retained via padding with black pixels at the borders where necessary to ensure the aspect ratio of the original is maintained.

### 2.2.3. Cropping

The raw images were first cropped to remove unnecessary parts of the limb that do not include the joints. For the hands, the bottom $\frac{1}{7}$ of the image was removed. This mostly removed the beginnings of the ulna and radius bones and part of the wrist. For the feet, the bottom ¼ of the image was removed. This did not remove any of the joints involved in scoring as the toes are found nearer the top of the feet.

*2.2.4. Noise removal and increasing image contrast*

The contrast of the images was then increased using CLAHE with a clip limit of 2 and the default grid size of (8,8). This was done to address the issue that some images were almost too dark or too light to be seen clearly. Fig. 2 show the importance of improving contrast.

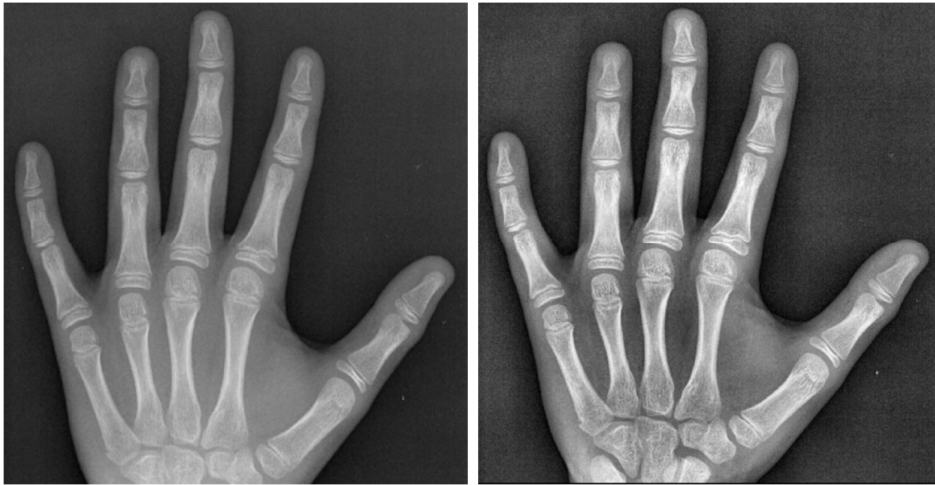

**Fig. 2.** Randomly selected before and after CLAHE processing images for side by side comparison.

*2.3. Joint Segmentation and Identification*

Apart from the joints themselves, the rest of the image is unnecessary information. As such, by first extracting only the joint images and using them to train a model to predict their scores would save computation and improve accuracy. Hence, a two-step method was used: U-Net background Masking, and YOLOv3.

*2.3.1. Mask Extraction Algorithm*

Before the U-Net could be trained and used for background removal, an algorithm was generated and applied on the images first. This included 3 main steps: 1) Getting entropy of the image, 2) applying Otsu thresholding, and, 3) further removal of mask noise.

Entropy is a measure of the amount of randomness in an image [17]. As there is variation in pixel values between the background and the skin and bones of the limb, this textural feature can be obtained as the entropy of the areas across the image. Entropy can be defined as follows:

$$H = -\sum_{i=0}^{n} p(x_i) \log_2 p(x_i) \tag{1}$$

where $n$ is the number of discrete levels of pixel values within a region of 37 x 37 pixels, and $p(x_i)$ is the probability of a pixel belong to the discrete level $i$, which is just the proportion of pixels that are in that level.

After obtaining the entropy across the image, Otsu thresholding [18] is applied to differentiate between background and limb. An intensity threshold level is supplied by minimizing the intensity variance within each class. This is defined as minimizing:

$$\sigma_\omega^2(t) = \omega_0(t)\sigma_0^2(t) + \omega_1(t)\sigma_1^2(t) \qquad (2)$$

where $\omega_0(t)$ and $\omega_1(t)$ are the probabilities of a pixel being in the two classes when separated by the threshold $t$. $\sigma_0^2$ and $\sigma_1^2$ are the intensity variances of the two classes.

The mask obtained from applying the threshold mostly have some amount of noise due to the varying and random X-ray photons detected in the background of the raw image. This method is adapted from a paper which only used flood filling [19], but it was found to be unsatisfactory. Hence, additional steps, such as contour identification and filling, and flood filling from multiple origins, were taken to further remove this noise.

Contours, the borders of a region with the same intensity, were first identified and filled with white pixels. Regions with sizes smaller than a threshold of 1% of the total white area were then removed by applying a layer of black pixels along its boundary. This mostly cleared any small regions of noise in the background.

Flood filling was then used to remove any noise found within the limb. The mask image was flood filled from the four corners in case any corner might have some noise remaining which could prevent the flood filling from working. These steps resulted in a mostly full and cleaned mask of the limb. The mask obtained was then applied onto the cropped image to remove all the noise from the background.

### 2.3.2. U-Net Masking

To achieve robustness against the background noise, the effectiveness of the nonuniform background noise removal step had to be improved. It is almost impossible to develop a robust unsupervised algorithm to remove noises which can come in various forms. As such, a deep learning model was used to learn how to generate masks of the limbs to remove noise in the background. The U-Net architecture was selected for this model. The traditional U-Net has been commonly used for semantic image segmentation especially for biomedical applications. It has been modified from a conventional Fully Convolutional Network (FCN) [20] so that it is able to perform well on medical images [21]. It is implemented like an encoder-decoder network but contains skip connections [21]. These skip connections create a link between layers and others that are deeper in the network. Unlike the classical encoder-decoder network, the output space mapping depends on both the latent space and the input space instead of only the latent space.

The specific architecture chosen was a lightweight U-Net architecture which was constructed for bones segmentation [13], but in this project, it was used for the purpose of background masking. Here, the number of down and up-sampling operations were adjusted to achieve higher performance in radiographic image segmentation [13]. Additionally, a multi-scale block (MSB) structure was used to do feature extraction. The MSB utilizes filters of different kernel sizes to deal with features at various scales [13].

Masks that were used to train the U-Net were first obtained using the Mask Extraction algorithm. From there, good mask outputs which match the limb well and do not contain noise from the background were selected by eye. A total of 296 hand masks and 238 feet masks were selected for training. These were then used to train 2 separate U-Net models, one for each type of limb. Fig. 3 shows examples of good masks that were chosen from the Mask Extraction algorithm.

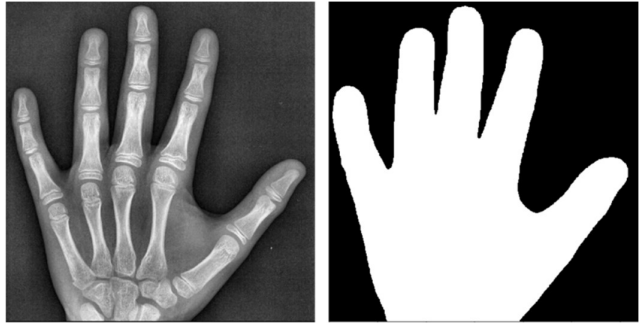

**Fig. 3.** Examples of good masks with their corresponding original images.

The optimiser used was the Adam optimizer with a learning rate set as 0.0001. The batch size was set at 16, and training ran for 200 epochs. Early stopping was used to prevent overfitting of the model. The loss function selected was binary cross entropy since a mask pixel can only either be 0 or 1. This is defined mathematically by:

$$BCE\ Loss = -\frac{1}{N}\sum_{i=1}^{N} y_i \cdot \log \hat{y}_i + (1 - y_i) \cdot \log (1 - \hat{y}_i) \qquad (3)$$

where $\hat{y}_i$ is the value of the $i$-th pixel from the predicted output, $y_i$ is its corresponding target value, and $N$ is the number of pixels in total.

### 2.3.3. YOLOv3 and Joint Identification

YOLOv3 is used for fast object detection where multiple bounding boxes would be predicted at the same time together with their class probabilities from one full image directly by a single network [14]. It outperforms Faster R-CNN (FRCNN) in terms of fewer background errors [14]. YOLOv3 is also generalizable as it was able to learn broader representations of images. Hence, it can be applied onto medical images such as radiographs in this case.

All selected images were annotated via the use of an open-source graphical image annotation tool called labelImg [22]. Since all the joints are of varying sizes, all boundary boxes had to be manually drawn individually.

The hands and feet images were trained separately. Since both the hands and feet contain the same number of classes, PIP (Proximal Interphalangeal) and MCP (Metacarpophalangeal) for hands, PIP and MTP (Metatarsal-Phalangeal) for feet, the parameters for training these 2 sets of data were the same. The configurations that were changed from the default settings are as follows:

1. max_batches = 4000
2. steps = 3200, 3600
3. classes = 2
4. filters = 21 for the end of each convolution blocks - layer 82, 94, and 106

However, the YOLOv3 models were not trained from scratch. A pre-trained weight on Common Object in Context (COCO) dataset, which could predict 80 classes were used [14]. The YOLOv3 trained models could now detect the ROIs and identify the type of joint - whether it is PIP or MCP for hand images, and whether it is PIP or MTP for foot images, by keeping the threshold at 0.5 confidence level even though most of the joints were classified to be around 95 –100% confident. If more than 10 or 6 joints were identified for the hands and feet respectively, the top 10 or 6 joints identified based on their confidence levels were picked (Fig. 4).

However, the predictions do not provide information of which PIP, MCP/MTP the joints belonged to. To determine which fingers and toes, such as index, middle, etc, a simple algorithm was developed to help in the identification. The algorithm was conducted based on the detected joints' positions along the horizontal axis. A backup algorithm was also created to determine joint type and location if the YOLOv3 model predicted the number of each joint type wrongly.

In the case where the model detects fewer than the required number of joints (10 for hands, and 6 for feet), the joint types identified by the model would not be utilized and their locations cannot be determined accurately. As such, this information would be unavailable during prediction of the scores for these joints. Consequently, the scores cannot be tagged to the exact joint.

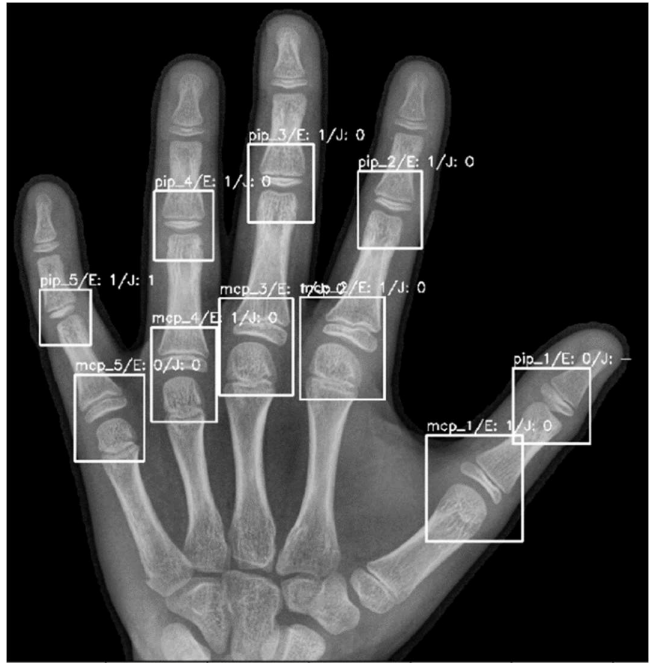

**Fig. 4.** Example of a patient's hand image with their ROI joints identified.

2.4. *Joint Scoring*

2.4.1. *Train, Validation, and Test Splits*

All train-test splits were conducted with test size equals to 10% of the entire data with a random state of 42. These test sets were not touched during training and were only used for final predictions and accuracy evaluations. During training, a validation split of 10% of the training sets was used.

2.4.2. *Metric for evaluating the model performance*

The joint scores for both erosion and narrowing take discrete integer values. As such, they can be deemed as separate classes. This means that joint scoring becomes a classification problem as compared to initially using RMSE for regression. However, it should be noted that these classes should be considered ordinal in nature.

The metric used to measure the performance of the models tested thus had to change. A modified accuracy score, the ±1 balanced accuracy, was used. To understand this, balanced accuracy is first defined as:

$$Balanced\ Acc = \frac{1}{N}\sum_{i=0}^{N-1}\left(\frac{Number\ of\ Correctly\ Predicted\ Class\ i}{Total\ Number\ of\ Class\ i}\right) \quad (4)$$

where $N$ is the total number of classes.

The ±1 balanced accuracy is similar to balanced accuracy just that samples which were predicted as a neighbouring class to its actual class (off by one class) would be considered as correctly predicted. This helps to account for the ordinality of the classes as well and an off-by-one classification is still medically acceptable [11]. This is also a suitable metric for a test set with an imbalanced distribution of the classes. In this case, a prediction of all samples as class-0 could achieve a high percentage accuracy. Hence, a better measure of performance would be how accurate the model is at predicting samples amongst each class. Using a balanced accuracy achieves this.

*2.4.3. VGG16 with Transfer Learning (TL)*

VGG16 is a vision model with CNN architecture that performs well in image classification [22]. Transfer learning refers to applying previously learnt knowledge from other tasks onto new but related tasks. Unfortunately, the pretrained weights for VGG16 were not trained on any radiographic images. As such, the model had to be trained on the radiographs using transfer learning.

The original VGG16 architecture was used but the convolution layers were frozen, leaving only the weights on the fully connected dense layers to be trainable. This was to prevent any changes in the pre-trained weights so the model could be tweaked to suit radiographic images.

*2.4.4. Ordinal Class Encoding*

Ordinal class encoding is a special encoding that is similar to multi-label classification encoding. It is as though the higher numbered classes have to have the labels of all the classes lower than them and one more additional label. This suggests that each class is a subset of the classes lower than it. For example, the encoding for classes 0 – 2 would be as follows:

Class 0: [1, 0, 0]
Class 1: [1, 1, 0]
Class 2: [1, 1, 1]

*2.4.5. Loss and activation Functions*

When using ordinal class encoding for the training of joints, the loss function needs to be binary cross entropy (BCE) with a 'sigmoid' final activation. The sigmoid function allows for the multi-labels to work. BCE is used here because this classification is done as multiple binary classifications. For example, the first binary classification would be between the 0 class samples and the non-0 class samples. The next binary classification would be between the 1 class samples and the non-1 class samples. This continues with all the other classes. In this way, the multi-labels and BCE will take the class ordering into account during training.

*2.4.6. Under-sampling*

Under-sampling refers to using fewer samples of a class for training. In this case, the number of joints with 0s was reduced to an amount close to the class that had the next highest sample size.

*2.4.7. Training*

A total of 4 models were trained. Two separate models for the erosion and narrowing for the joints in the hands, and 2 separate models for erosion and narrowing of the feet joints. All models were trained for 250 epochs with learning rate of 0.0001. As the convolution layers were frozen, the number of trainable parameters was reduced to:

- Total parameters: 15,113,049
- Trainable parameters: 396,815
- Non-trainable parameters: 14,716,234

## 2.5. Full Pipeline

These models were then integrated into the end of the pipeline that includes the stages of pre-processing, masking, joint identification and extraction. They are meant to perform the final step in the pipeline: joint score prediction. The raw radiographs would be passed into this pipeline as input while the individual joint scores and overall scores would be outputted together with the labelled images. See Fig 5 for the visualization of the full pipeline.

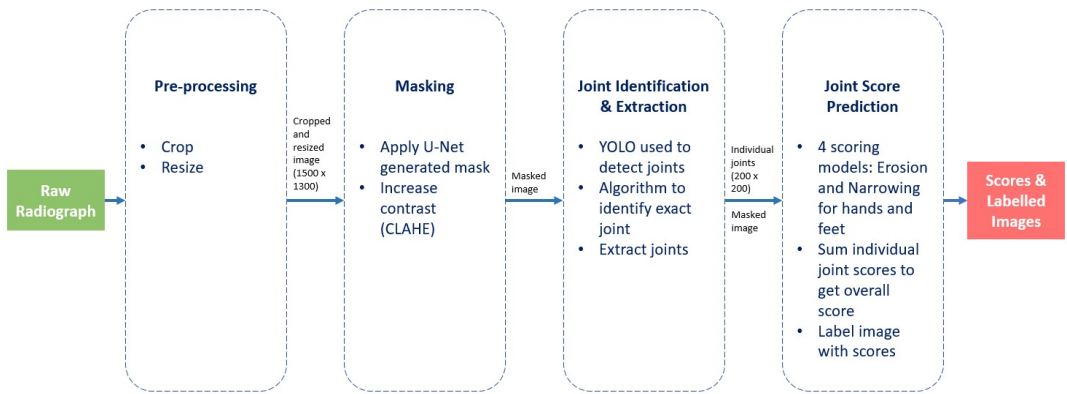

**Fig 5.** Full pipeline.

## 3. Results

Since the joint detection and identification step using the lightweight U-Net and YOLOv3 respectively was proved to have an accuracy of 99.991%, the accuracy of the full pipeline with these models integrated at the joint scores prediction step can be approximated as the same accuracies shown in Table 1, which shows the joint-wise ±1 balanced accuracy for each prediction model. See Fig. 6 for the models' confusion matrix.

**Table 1** Joint-wise ±1 balanced accuracy of each prediction model.

| Model | Balanced Test Accuracy (%) | ±1 Balanced Accuracy (%) |
| --- | --- | --- |
| Narrowing (Hands) | 82.07 | 97.30 |
| Narrowing (Feet) | 83.76 | 91.51 |
| Erosion (Hands) | 78.77 | 94.63 |
| Erosion (Feet) | 57.83 | 92.49 |

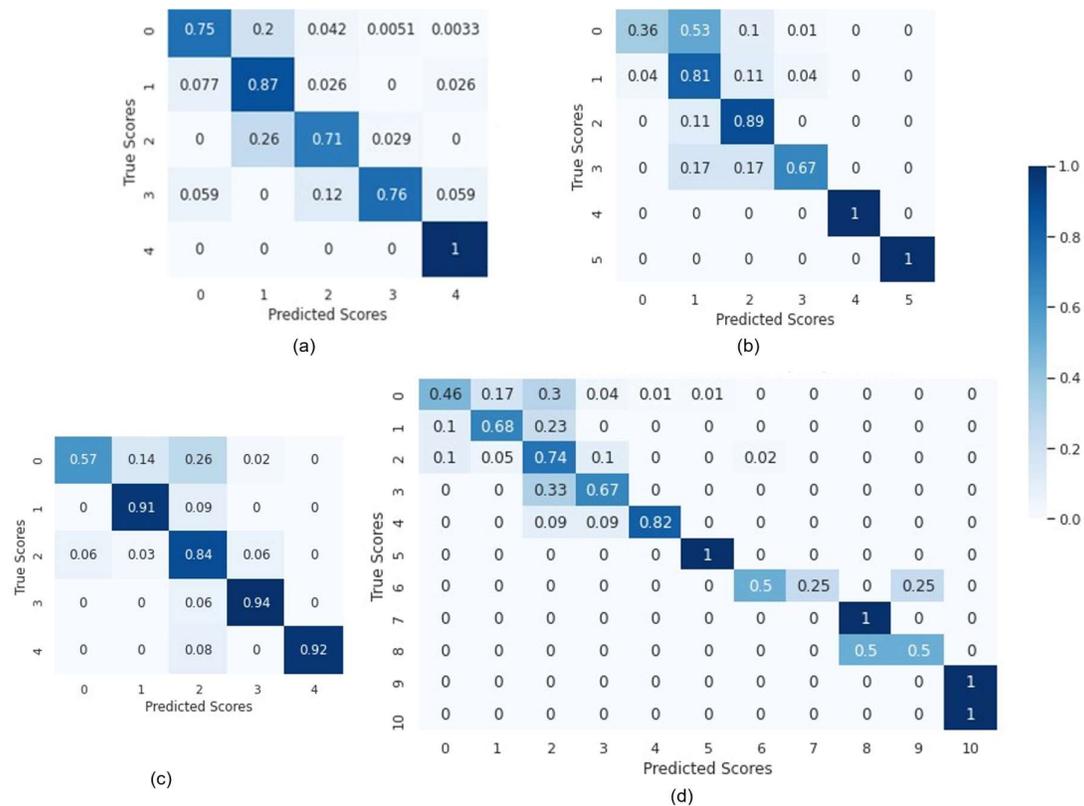

**Fig. 6.** Confusion matrices for the +- balanced accuracies on the test set of (a) joint narrowing (Hands) model; (b) Joint erosion (Hands) model; (c) Joint narrowing (Feet) model; (d) Joint erosion (Feet) model.

## 4. Discussion

### 4.1. Summary of Results

The ±1 balanced accuracy of the 4 models showed great potential in achieving industry standard reliability. However, the balanced accuracy without ±1 tolerance is not as promising as it ranges from 57.83% to 83.76%. This suggests that if being exact is important, this approach does not do very well in achieving that. However, since most methods to determine RA severity relies on subjective measures, it shows that such a tool would still be able to help by getting a gauge of the disease progression. It might not be suitable to rely on this method to get an objective measure with high accuracy for standardized use in research. On the other hand, by allowing physicians to save time by automatically predicting joint damage at a high accuracy, it can still be used as part of the process in determining RA severity rather than just on its own.

*4.2. Possible Impact*

Currently, the method can be used as an automatic tool to get a gauge on the degree of joint damage. This might not be able to serve as a standardization for clinical or research use, but it can still help in determining RA severity. Subjective measures are still used in the clinical setting due to limited time and resources. As such, this method can provide a highly accurate and quick measure of the amount of joint damage which is indicative of RA severity.

For more extensive use, the model needs to be able to be reliable enough for use in the industry where patients' wellbeing is involved. Hence, the standards for the model must be extremely high. The approach proposed here and the performance it achieved shows great promise that it can reach this high standard. Once it is achieved, the solution would have a great impact on RA patient care through better disease management. This would be due to saving time and effort from manual scoring and providing an objective measure of the degree of damage for clinical diagnoses. It would also aid research that requires the use of objective joint damage scores.

*4.3. Future Work/ Limitations*

To achieve a standard high enough for use, some possible further improvements to overcome certain limitations identified could be implemented.

Firstly, improvements on the approach could be made. Bone segmentation could be used to remove the skin tissue in the radiograph. This could help improve the joint detection rate and increase prediction accuracy as there is less unnecessary information from the skin. Another improvement could be to do binary classification of the 0 class and non-0 classes before differentiating between the non-0 classes. This could help with the imbalance in the distribution of classes where there are a lot of 0 class joints. Computer vision could also be used to get the physical characteristics of the joint spaces such as distance measures for narrowing scores or the contours of the joint.

Apart from adjusting the method, accuracy of the prediction model can also be improved by obtaining more data samples for training. This could be done by getting data from hospitals and laboratories. Hopefully, with these improvements, the model will be able to achieve a high level of accuracy and hence, reliability, so that it can be deployed, and have its benefits realized.

## 5. Conclusion

Using a joint segmentation approach and training joint score prediction models with ordinal class encoding, under-sampling and transfer learning, the joint wised ±1 balanced accuracies ranging from 91.51% to 97.30% were achieved.

## Acknowledgements

The Datasets used for the analyses described in this manuscript (or publication) were contributed by University of Alabama at Birmingham. They were obtained as part of the RA2-DREAM Challenge: Automated Scoring of Radiographic Damage in Rheumatoid Arthritis through Synapse ID [syn20545111]